\documentclass[prd,superscriptaddress,a4paper,showpacs,showkeys,11pt,nofootinbib]{revtex4}
\usepackage{graphicx}
\usepackage{graphics}
\usepackage{dcolumn}
\usepackage{bm}
\usepackage{multirow}
\usepackage{tabularx}
\usepackage{hyperref}
\usepackage{commath}
\usepackage{epstopdf}
\usepackage[T1]{fontenc}
\usepackage[latin9]{inputenc}
\usepackage{geometry}
\usepackage{soul}
\usepackage{amsmath,amssymb,amsfonts}

\usepackage{color}

\def\be{\begin{equation}}
\def\ee{\end{equation}}

\providecommand{\ee}{e$^+$e$^-$}

\makeatother

\begin{document}

\begin{flushright}
MS-TP-22-24
\end{flushright}

%
%
\title{Exclusive doubly charged Higgs boson pair production in $pp$ collisions at the LHC }


\author{Laura {\sc Duarte}}
\email{l.duarte@unesp.br}
\affiliation{International Institute of Physics, Universidade Federal do Rio Grande do Norte,
Campus Universitario, Lagoa Nova, Natal-RN 59078-970, Brazil }

\author{Victor P. {\sc Gon\c{c}alves}}
\email{barros@ufpel.edu.br}
\affiliation{Institut f\"ur Theoretische Physik, Westf\"alische Wilhelms-Universit\"at M\"unster,
Wilhelm-Klemm-Stra\ss e 9, D-48149 M\"unster, Germany}
\affiliation{Institute of Modern Physics, Chinese Academy of Sciences,
  Lanzhou 730000, China}
\affiliation{Institute of Physics and Mathematics, Federal University of Pelotas, \\
  Postal Code 354,  96010-900, Pelotas, RS, Brazil}

\author{Daniel E. {\sc Martins}}
\email{dan.ernani@gmail.com}
\affiliation{Institute of Physics and Mathematics, Federal University of Pelotas, \\
  Postal Code 354, 96010-900, Pelotas, RS, Brazil}
  \affiliation{The Henryk Niewodniczanski Institute of Nuclear Physics (IFJ) - Polish Academy of Sciences (PAN), \\
    31-342, Krakow, Poland}

\author{T\'essio B. de {\sc Melo}}
\email{tessiomelo@gmail.com}
\affiliation{Institute of Physics and Mathematics, Federal University of Pelotas, \\
  Postal Code 354,  96010-900, Pelotas, RS, Brazil}

\author{Farinaldo S. {\sc Queiroz}}
\email{farinaldo.queiroz@ufrn.br}
\affiliation{International Institute of Physics, Universidade Federal do Rio Grande do Norte,
Campus Universitario, Lagoa Nova, Natal-RN 59078-970, Brazil }
\affiliation{Departamento de Fisica, Universidade Federal do Rio Grande do Norte, 59078-970, Natal, RN, Brazil}
\affiliation{Millennium Institute for SubAtomic Physics at the High-energy frontIeR, SAPHIR, Chile}


\begin{abstract}
The type II seesaw model predicts the existence of a doubly charged Higgs boson ($H ^{\pm\pm}$), which can be produced through the Drell - Yan process and photon -- induced interactions at hadronic colliders. In recent years, such reactions have been largely investigated in inclusive processes, where both incident protons breakup and a large number of particles is produced in addition to the doubly charged Higgs pair. In this paper, we investigate, for the first time, the possibility of searching for $H ^{\pm\pm}$ in exclusive processes, which are characterized by  two rapidity gaps and two intact very forward
protons in final state. We estimate the associated cross section making use of the exclusive
  nature of the final state, together with the use of timing information
  provided by forward proton detectors. Moreover, the background is estimated considering distinct amounts of pile-up. Our results indicate that the exclusive doubly charged Higgs pair production is a promising way to search for signatures of the type II seesaw mechanism and to obtain lower mass bounds on $H ^{\pm\pm}$.
\end{abstract}



\keywords{Doubly charged Higgs, exclusive processes, proton-proton collisions}

\maketitle

\section{Introduction}

Many beyond Standard Model (SM) scenarios predict the existence of a scalar particle carrying two units of electric charge, especially those related to neutrino masses. One of the most popular scenarios featuring this kind of particle is the type II seesaw mechanism \cite{Schechter:1980gr, Cheng:1980qt, Arhrib:2011uy, Ashanujjaman:2021txz}, in which a $SU(2) _L$ scalar triplet $\Delta = (H ^{++}, H ^+, H ^0)$ accounts for the tiny neutrino masses after its neutral component develop a vacuum expectation value (VEV). Accompanying the neutral scalar inside the triplet, there is the doubly charged scalar $H ^{++}$ (along with a singly charged scalar as well). Doubly charged scalars are also present in left-right symmetric models \cite{Pati:1974yy, Mohapatra:1974gc, Senjanovic:1975rk} (which were, in fact, the first models to propose the existence of these particles), coming from either a $SU(2) _L$ or a $SU(2) _R$ triplet. 
They can also arise from other scalar multiplet representations in models such as radiative seesaw models \cite{Zee:1985id, Babu:1988ki}, 3-3-1 models \cite{CiezaMontalvo:2006zt, Alves:2011kc, Machado:2018sfh, Ferreira:2019qpf}, Georgi-Machacek \cite{Georgi:1985nv, Chanowitz:1985ug, Chiang:2018cgb} and little Higgs models \cite{Arkani-Hamed:2002iiv, Arkani-Hamed:2002ikv, Hektor:2007uu}.

Several searches for doubly charged scalar particles have been performed recently, specially in high-energy colliders \cite{ATLAS:2012hi, CMS:2012dun, ATLAS:2014kca, CMS:2014mra, ATLAS:2017xqs, CMS:2017fhs, ATLAS:2018ceg, ATLAS:2022pbd}. From the non-observation of any excess compared to the SM background, limits on the production cross section of the $H ^{++}$ have been placed, which can be translated into bounds on the $H ^{++}$ mass, 
depending on the assumptions about its decays modes. Since in many scenarios neutrino masses are generated via the coupling between the SM lepton doublet and the scalar multiplet that hosts the $H ^{++}$, a reasonable assumption is that $H ^{++}$ is able to decay into SM leptons. In particular, the decay into same-sign dileptons provides a smoking gun signature for a doubly charged particle and, for this reason, this signature has been used in most of the experimental analysis. { The 
most stringent limits to date come from the searches by the ATLAS Collaboration obtained with 139 fb$^{-1}$ data collected at 13 TeV \cite{ATLAS:2022pbd}. 
 Assuming that 
 decays to each of the $ee$, $e\mu$, $\mu\mu$, $\mu\tau$, $e\tau$, $\tau\tau$  final states
are equally probable, the lower limit on $H^{\pm\pm}$ mass is around 1080 GeV, at $95 \%$ C.L. within
the type II seesaw model.}

\begin{figure}[t]
\includegraphics[scale=0.5]{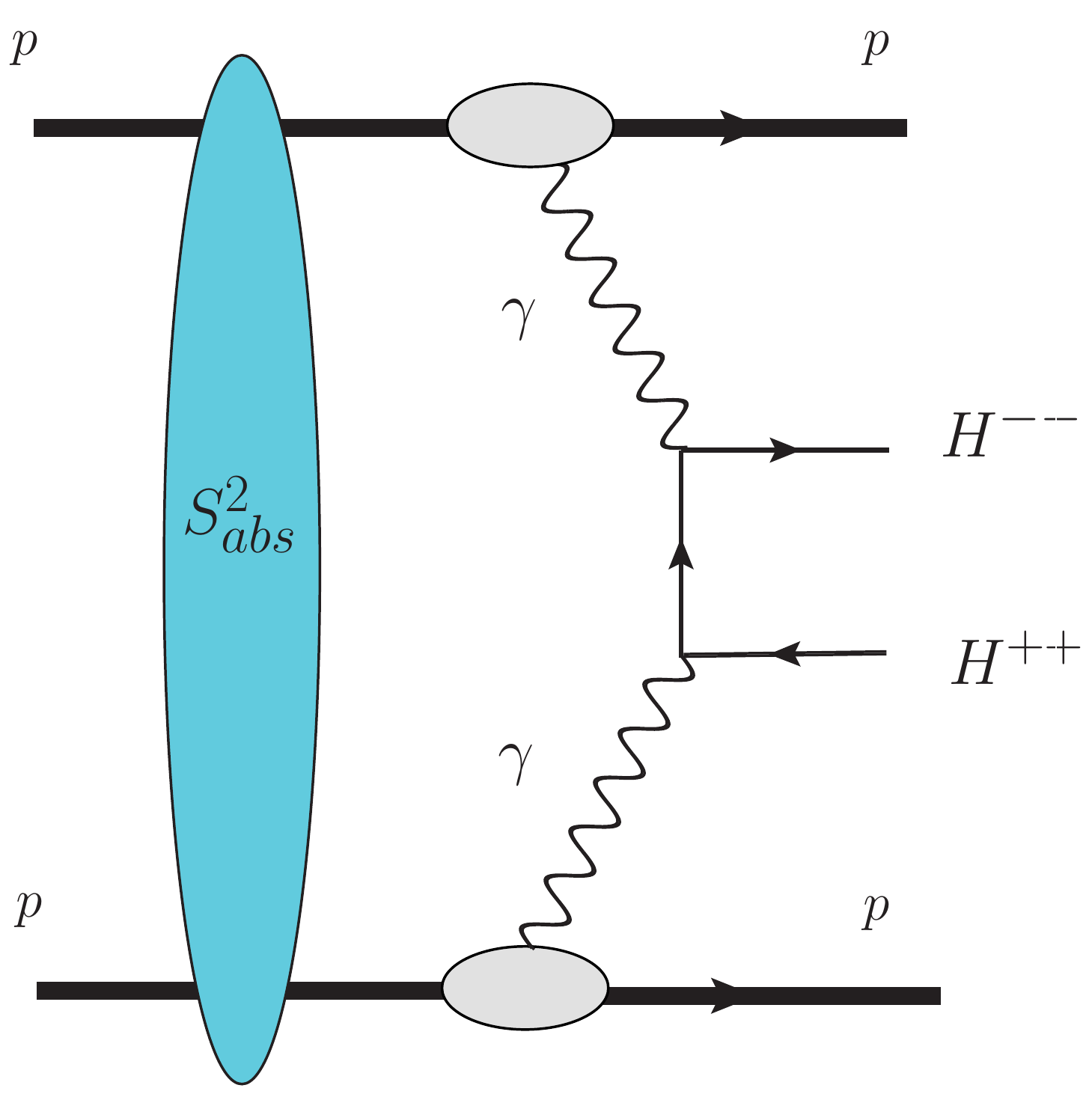}
\caption{Diagram for the exclusive doubly charged Higgs pair production in $pp$ collisions at the LHC.}
\label{Fig:diagram}
\end{figure}

The current  lower limit on the mass of the doubly charged Higgs in hadronic colliders have been mainly obtained by considering that the $H^{+ +} H^{- -}$ pair can be produced via the Drell - Yan (DY) process as well as by photon - photon  and gluon - gluon interactions. As demonstrated e.g. in Ref.  \cite{Fuks:2019clu}, photon - induced processes are suppressed by about an order of magnitude with respect to the DY production, and the gluon fusion is subdominant compared to the $\gamma \gamma$ fusion. The DY production is an inclusive process, characterized by the fragmentation of both incident protons and a large number of particles is produced in addition to the doubly charged Higgs pair.
On the other hand, in $\gamma \gamma$ interactions, the processes can be classified as elastic, semi - elastic or inelastic, depending if both, only one or none of the incident protons remain intact (For previous studies see, e.g. Refs. \cite{Han:2007bk,Babu:2016rcr}). As a consequence,  the elastic events, represented in Fig. \ref{Fig:diagram},  although subdominant with respect to the other channels, has a final state characterized by the presence of two regions devoid of hadronic activity, called rapidity gaps, separating the intact very forward
protons from the central massive object. Such very clean final state is a characteristic of exclusive processes, which are generated by the exchange of  color singlet objects (e.g. a photons or a Pomeron) \cite{pomeron}. Such distinct topology of exclusive processes was  not explored in previous studies and will be considered, for the first time, in the present paper.  
In our analysis we will consider that 
these events can be collected using the forward proton
detectors (FPD) such as the ATLAS Forward Proton detector (AFP)
\cite{Adamczyk:2015cjy,Tasevsky:2015xya} and Precision Proton Spectrometer
(CT-PPS) \cite{Albrow:2014lrm} that are installed symmetrically around
the interaction point at a distance of roughly 210~m from the interaction point. 
In particular, we will assume that $m(H^{\pm \pm})´ \ge 350$ GeV and that the doubly charged Higgs decays into same - signal dileptons. The large invariant mass of the produced system implies that the intact protons
in the final state can be tagged by FPDs. Consequently, such events can, in principle, be separated and be used to
searching for the doubly charged Higgs and/or impose limits on its mass. In our analysis we will also consider the 
presence of  pile-up events, which is a reality in current and future LHC runs, and arise due  to the presence of extra $pp$ interactions per bunch crossing in high luminosity $pp$ collisions at the
LHC. These extra interactions generates additional tracks  increase the background stemming from the inelastic doubly charged Higgs pair produced in a
different primary vertex and, in general, destroy the signature associated to two
rapidity gaps. Following Refs. \cite{Goncalves:2020saa,Martins:2022dfg}, such  background will be suppressed by considering the time-of-flight (ToF) detectors \cite{Harland-Lang:2018hmi,Tasevsky:2014cpa,ToFperformance,ToFPUBNote}. 
As we will demonstrate below, the exclusive $H^{+ +}H^{- -}$ production is a promising alternative to searching for signals of the type II seesaw mechanism.  { It is important to emphasize that over the last years the searching  of BSM physics in exclusive processes has  been proposed by several groups (see, e.g., Refs.   \cite{Harland-Lang:2018hmi,Ohnemus:1993qw,Harland-Lang:2011mlc,Fichet:2013gsa,Goncalves:2014jea,Goncalves:2015oua,Lebiedowicz:2015cea,Alva:2014gxa,Lebiedowicz:2016lmn,Baldenegro:2017aen, 
Ghosh:2017pxl,KumarGhosh:2018bli,Goncalves:2020bqi}) and that the experimental analysis performed by the ATLAS and CMS Collaborations of events with proton tagging          already have imposed important constraints on several scenarios for New Physics \cite{CMS:2018uvs,ATLAS:2020mve,TOTEMCollaboration:2021xam,CMS:2022dmc,RibeiroLopes:2022rgp}. Our results indicate that an experimental analysis is feasible and that the associated data will be very useful to improve our understanding of the type II seesaw mechanism. }

This paper is organized as follows. In the next section we present a brief
review of the formalism for the doubly charged Higgs interactions and for the treatment of the $H^{\pm \pm}$ pair production in photon-induced interactions in $pp$ collisions.  In
Section~\ref{sec:results} we discuss the backgrounds considered in our analysis as well as the exclusivity cuts. In addition, we present our predictions for the
invariant mass, transverse momentum and rapidity distributions as well as for the total
cross sections for the $H^{\pm \pm}$ pair production in $\gamma \gamma$ interactions. Finally, in Section~\ref{sec:sum} we summarize our main
conclusions.

\section{Formalism}
\label{sec:form}

In this section we describe the $H ^{++}$ basic features and the relevant interactions for its production and decay at the LHC.
Since doubly charged scalar bosons appear in several extensions of the SM \cite{Basso:2015pka, Zee:1985id, Georgi:1985nv, Kumericki:2012bh}, here we will first focus on the case of the Higgs Triplet Model, which is a minimal realization of the type II seesaw mechanism, and later extend to more general $H ^{++}$ particles, keeping some of the key features from this prototypical case.

In the Higgs Triplet Model, the SM scalar sector is extended by adding a weak scalar triplet $\Delta$, 
\begin{equation}
\label{eq:triplet}
\Delta (1, 2, 1) = \begin{pmatrix} \Delta ^+ / \sqrt{2} & \Delta ^{++} \\ \Delta ^0 & - \Delta ^+ / \sqrt{2} \end{pmatrix} ,
\end{equation}
where the numbers in parentheses denote the representation under the $SU(3)_C\times SU(2)_L\times U(1)_Y$ gauge group of the SM \footnote{We are using the convention $Q = T _3 + Y$ for the electric charge and hypercharge.}.
{The most general renormalizable scalar potential is given by
\begin{equation}\label{potential}
\begin{split}
V = & - m _\Phi ^2 \Phi ^\dagger \Phi + \frac{\lambda}{4}(\Phi ^\dagger \Phi) ^2 + m _\Delta ^2 \text{Tr} \Delta ^\dagger \Delta + ( \mu \Phi ^T i \sigma ^2 \Delta ^\dagger \Phi + \text{h.c.}) \\
& + \lambda_1 \Phi ^\dagger \Phi \text{Tr} \Delta ^\dagger \Delta + \lambda _2(\text{Tr} \Delta ^\dagger \Delta) ^2 + \lambda _3 \text{Tr} (\Delta ^\dagger \Delta) ^2 + \lambda _4 \Phi ^\dagger \Delta \Delta ^\dagger \Phi ,
\end{split}
\end{equation}
where $\Phi$ is the SM-like scalar doublet, $m _\Phi ^2$, $m _\Delta ^2$ and $\mu$ are mass parameters, $\lambda$ and $\lambda_i$
$(i = 1, ..., 4)$ are independent dimensionless couplings. After spontaneous symmetry breaking, both scalars $\Phi ^0$ and $\Delta ^0$ acquire VEVs, denoted by $\langle \Phi ^0 \rangle = v /\sqrt{2}$ and $\langle \Delta ^0 \rangle = v _\Delta / \sqrt{2}$. From the two minimization conditions of the scalar potential, we can express $m _\Phi ^2$ and $m _\Delta ^2$ in terms of other parameters as,
\begin{equation}
\begin{split}
& m _\Phi ^2 = \frac{\lambda}{4} v ^2 + \frac{\lambda _1 + \lambda _4}{2}v _\Delta ^2 - \sqrt{2} \mu v _\Delta , \\
m _\Delta ^2 & = - \frac{(\lambda _1 + \lambda _2)}{2} v ^2 - (\lambda _2 + \lambda _3) v _\Delta ^2 + \frac{\mu v ^2}{\sqrt{2} v _\Delta} .
\end{split}
\end{equation}
The neutral components of $\Phi$ and $\Delta$ can be parametrized as, 
\begin{equation}
\begin{split}
\Phi ^0 & = (\rho _1 + v + i \eta _1)/ \sqrt{2} \\
\Delta ^0 & = (\rho _2 + v _\Delta + i \eta _2)/ \sqrt{2} ,
\end{split}
\end{equation}
and after the electroweak symmetry breaking, the mixing among the scalar fields lead to several Higgs bosons. The gauge eigenstates can be rotated to obtain the mass eigenstates as in the following:
\begin{equation*}
\begin{split}
\begin{pmatrix} h \\ H \end{pmatrix} = R(\alpha) \begin{pmatrix} \rho _1 \\ \rho _2 \end{pmatrix} , \quad \begin{pmatrix} G \\ A \end{pmatrix} = R(\beta) \begin{pmatrix} \eta _1 \\ \eta _2 \end{pmatrix} , \quad \begin{pmatrix} G ^+ \\ H ^+ \end{pmatrix} = R(\beta ^\prime) \begin{pmatrix} \phi _1 ^+ \\ \Delta ^+ \end{pmatrix} ,
\end{split}
\end{equation*}
where $R(\theta)$ is the usual rotation matrix,
\begin{equation*}
R(\theta) = \begin{pmatrix} \cos \theta & \sin \theta \\ - \sin \theta & \cos \theta \end{pmatrix} ,
\end{equation*}
and $\alpha$, $\beta$ and $\beta^\prime$ are the rotation angles in the CP-even, CP-odd and singly-charged Higgs sectors. 
$h$ and $H$ are the CP-even neutral Higgs with masses $m _h$ and $m _H$ given by,
\begin{equation}
\begin{split}
m _h & = \frac{1}{2}(a + c - \sqrt{(a - c) ^2 + 4 b ^2}), \\
m _H & = \frac{1}{2}(a + c + \sqrt{(a - c) ^2 + 4 b ^2}),
\end{split}
\end{equation}
where $ a = \frac{\lambda}{2} v ^2$, $ b = - \sqrt{2} \mu v + (\lambda _1 + \lambda _4) v v _\Delta$ and $c = \frac{\mu v ^2}{\sqrt{2} v _\Delta} + 2 (\lambda _2 + \lambda _3) v _\Delta ^2$. The lightest of the neutral scalar mass eigenstates, $h$, is identified with the $125$ GeV SM-like Higgs boson.
The field $A$ is the CP-odd neutral scalar with mass $m _A$,
\begin{equation}
m _A ^2 = \frac{\mu (v ^2 + 4 v _\Delta ^2)}{\sqrt{2} v _\Delta} ,
\end{equation}
while $G ^+$ and $G ^0$ are the Nambu-Goldstone bosons which become the longitudinal modes of $Z$ and $W ^+$, respectively. $H ^+$ is the singly-charged Higgs boson, whose mass $m _{H ^+}$ is given by,
\begin{equation}
m _{H ^+} = \frac{(v ^2 + v _\Delta ^2)(2 \sqrt{2} \mu - \lambda _4 v_\Delta)}{4 v _\Delta}.
\end{equation}
Finally, the doubly-charged Higgs $H ^{++} (\equiv \Delta^{\pm\pm})$ comes solely from the triplet, and has a mass:
\begin{equation}
m _{H ^{++}} ^2 = \frac{\mu v ^2}{\sqrt{2} v _\Delta} - \frac{1}{2} ( \lambda _4 v ^2 + 2 \lambda _3 v _\Delta ^2 ) .
\end{equation}}

The triplet $ \Delta $ interacts with the SM leptons via the coupling,
\begin{equation}
\label{eq:yuk_lept_triplet}
\mathcal{L} _{Y} \supset - y \overline{L ^c} i \sigma ^2 \Delta L + h.c.
\end{equation}
where $y$ is the matrix of Yukawa coupling constants, $L = ( \nu _L \ e _L ) ^T $ is the SM lepton doublet, $\sigma ^2$ is the second Pauli matrix and the superscript $c$ denotes charge conjugation. By means of this interaction, neutrino masses are generated after $\Delta ^0$ develops its VEV, $v _\Delta$. 
{As the masses will be proportional to $v _\Delta$, it is clear that this VEV is required to be small in order to account for the tiny neutrino masses. On the other hand, a small $v _\Delta$ value lead to high triplet scalar masses. For values of the $\mu$ parameter around the electroweak scale, scalars with masses of hundreds of GeV to TeV are easily attainable. Although these extra scalars are very important if one wants to verify the Higgs Triplet Model, for the searches of the $H ^{++}$ that we will focus on this paper, they will play no role. Therefore, in what follows we will no longer discuss about them.}

The $H ^{++}$ in the Higgs Triplet Model further couples to gauge bosons and other scalars at tree level. The coupling to quarks, however, is forbbiden (note that the $Y = 1$ hypercharge that allows the coupling of $\Delta$ with the leptons in Eq. \eqref{eq:yuk_lept_triplet}, automatically forbids a coupling to the quark doublets).
Thus, the $H ^{++}$ has basically two decays modes: $H ^{\pm \pm} \rightarrow \ell _i ^\pm \ell _j ^\pm$ and $H ^{\pm\pm} \rightarrow W ^\pm W ^\pm$. The widths for these two body decays are given by \cite{Melfo:2011nx}:
\begin{equation}
\Gamma( H ^{\pm \pm} \to \ell _i ^\pm \ell _j ^\pm ) = \frac{m _{H ^{\pm \pm}}}{8 \pi ( 1 + \delta_{ij} )} \frac{ |\mathcal{M} _\nu ^{ij} | ^2}{v _\Delta ^2} ,
\end{equation}

\begin{equation}
\begin{split}
\Gamma \left( H ^{\pm \pm} \rightarrow W ^\pm W ^\pm \right) = \dfrac{g ^4 v ^2 _\Delta}{8 \pi m _{H ^{\pm \pm}}} \sqrt{1 - \dfrac{4 m ^2 _W}{m ^2 _{H ^{\pm \pm}}}} \left[2 + \left( \dfrac{m ^2 _{H ^{\pm \pm}}}{2 m _W ^2} - 1 \right) ^2 \right] ,
\end{split}
\end{equation}
where $\delta_{ij}$ is the Kronecker delta, $\mathcal{M}_{\nu}^{ij}$ are the elements of the neutrino mass matrix and $\ell_i^{\pm} = e^{\pm}, \mu^{\pm}, \tau^{\pm}$. 
From these expressions, it is clear that the branching ratios for $H ^{++}$ decays are very sensitive to $v _\Delta$. Most of the searches so far have been focused in the same sign dilepton decays $H ^{\pm \pm} \rightarrow \ell _i ^\pm \ell _j ^\pm$, which provides a clean distinctive signature. This amounts to assume that $v _\Delta \leq 10 ^{-4}$ GeV, so that this decay is dominant over the $H ^{\pm\pm} \rightarrow W ^\pm W ^\pm$ channel. Although this channel is more difficult to analyze due to the presence of neutrinos in the final state, and the consequent larger SM backgrounds, searches in this channel have been done \cite{ATLAS:2018ceg, ATLAS:2022pbd}. It should be mentioned that, if kinematically allowed, $H ^{++}$ can also have cascade decays to scalar and vector bosons, such as $H ^{\pm \pm} \to H ^\pm W ^\pm \to H ^0 W ^\pm W ^\pm$. However, these decays are much more challenging to probe experimentally.

Doubly-charged Higgs bosons can be pair produced at the LHC via the Drell-Yan (DY) process $q \bar{q} \to \gamma ^* / Z ^* \to H ^{\pm \pm} H ^{\mp \mp}$. The trilinear and quartic gauge interactions relevant for the calculation of the pair production come from the kinetic Lagrangian for $H ^{++}$, and can be written as,
\begin{equation}
\begin{split}
\label{eq:lag_kin}
\mathcal{L} _{\text{kin}} = & i \left[ 2 e A _\mu + \dfrac{g}{c _W} \left( 2 - Y - 2 s _W ^2 \right) Z _\mu \right] H ^{++} \partial ^\mu H ^{--} \\
& + \left[ 2 e A _\mu + \dfrac{g}{c _W} \left( 2 - Y - 2 s _W ^2 \right) Z _\mu \right] \left[ 2 e A ^\mu + \dfrac{g}{c _W}\left( 2 - Y - 2s _W ^2 \right) Z ^\mu \right] H ^{++} H ^{--} ,
\end{split}
\end{equation}
where $c _W = \cos \theta _W$, $s _W = \sin \theta _W$, with $\theta _W$ being the weak mixing angle. 
The $H ^{++}$ can also be produced in association with a $W$ boson. However, in the Higgs Triplet Model this production rate is highly suppressed, since it is proportional to the factor $(v _\Delta ^2 / m _W ^2)$. Production of $H ^{++}$ in association with $H ^-$ can occur at significant rates via the process $u \bar{d} \to W ^{+*} \to H^{++} H ^-$ (See e.g. Ref. \cite{Akeroyd:2005gt}). { However, the pair production of a doubly charged scalar points more precisely to the doubly charged scalar mass, which is a common particle in several model building endeavors}  \cite{CMS:2012dun,CMS:2017pet}. For these reasons, in their searches for doubly charged Higgs bosons, the ATLAS and CMS collaborations have focused on the pair production via DY process. In the next section we shall discuss a new way to search for the $H ^{++}$, employing elastic photon fusion production, combined with the tagging of the intact protons after the collision, which is complementary to the ATLAS and CMS searches and leads to excellent detection prospects.

In our analysis, we also consider doubly charged scalars arising from more general $SU(2) _L$ representations. We assume that most of the features of the $H ^{++}$ from the Higgs triplet model remain valid for these other cases, in particular concerning its production and decay modes. More specifically, our analysis includes:
\begin{itemize}
\item $H ^{++}$ from a $SU(2) _L$ singlet:
$\phi = H ^{++}$; ($T _3 = 0, Y = 2$);
\item $H ^{++}$ from a $SU(2) _L$ doublet:
$\phi = (H ^{++} \ H ^+)$; ($T _3 = 1/2, Y = 3/2$);
\item $H ^{++}$ from a $SU(2) _L$ triplet:
$\phi = (H ^{++} \ H ^+ \ H ^0)$; ($T _3 = 1, Y = 1$).
\end{itemize}
These representations arise quite commonly in a variety of models. For instance, in some versions of the 3-3-1 model \cite{Ferreira:2019qpf,Machado:2018sfh}, an scalar triplet emerges from an original $SU(3) _L$ scalar sextet, after this symmetry is spontaneously broken down to the SM gauge group. In the left-right symmetric model \cite{Borah:2016hqn, BhupalDev:2018tox}, besides the $H _L ^{++}$ from the $SU(2)_L$ triplet $\Delta _L$, there is also a companion $H _R ^{++}$ from the $SU(2) _R$ triplet $\Delta _R$ (which is however singlet under $SU(2) _L$). In radiative neutrino mass models \cite{Zee:1985id, Babu:1988ki}, the doubly charged Higgs boson is analogous to the singlet $H _R ^{++}$ of the left-right model. 
As the presence of singlets, doublets or triplets containing a doubly charged boson is very common in  several neutrino mass studies \cite{Arhrib:2009mz,Bandyopadhyay:2009xa,Toma:2013zsa,Vicente:2014wga,Fraser:2014yha,Biswas:2019ygr}, we will stick to these representations. It should be noted that other scalar multiplets, such as $SU(2) _L$ quadruplets or quintuplets, can also host a $H ^{++}$. In these cases, however, the coupling of $H ^{++}$ to the leptons would require the presence of extra vector-like fermions that mix with the SM leptons. For this reason, we will not consider these higher isospin representations here. It is important to emphasize that the gauge interactions for the DY pair production given in Eq. (\ref{eq:lag_kin}) remain valid for all the aforementioned representations (provided the appropriate hypercharge value is used), while the production via photon-photon fusion, discussed below, is insensitive to the multiplet representation the $H ^{++}$ originates from. In addition, we will assume that, similarly to the case of the Higgs Triplet Model in the regime of small $v _\Delta$, the $H ^{++}$ decays dominantly into same-sign dileptons, also for the singlet and doublet cases.

\subsection{Production via photon-photon fusion}

As discussed in the Introduction, our focus in this paper is on the analysis of the doubly charged Higgs production by elastic $\gamma \gamma$ interactions, as represented in Fig. \ref{Fig:diagram}.  The total cross section for these interactions can be factorized in terms of the equivalent flux of
photons into the hadron projectiles and the photon--photon cross section. In particular, the $H^{++} H^{--}$  production in photon--photon interactions is described by
\begin{eqnarray}
\sigma(h_1 h_2 \rightarrow h_1 \otimes H^{++} H^{--} \otimes h_2) = S^2_{abs}  \int dx_1 \int dx_2 \, \gamma^{el}_1(x_1) \cdot \gamma^{el}_2(x_2) \cdot \hat{\sigma}(\gamma \gamma \rightarrow H^{++} H^{--}) \,\,,
\label{fotfot}
\end{eqnarray}
where $\otimes$ represents the presence of a rapidity gap in the
final state, $S^2_{abs}$ is the absorptive factor (see below), $x$ is the fraction of the hadron energy carried by the photon
and $\gamma^{el}(x)$ is the elastic equivalent photon distribution of the hadron. 
The general expression for the elastic photon flux of the proton is given by \cite{kniehl} 
\begin{eqnarray}
\gamma^{el} (x) = - \frac{\alpha}{2\pi} \int_{-\infty}^{-\frac{m^2x^2}{1-x}} \frac{dt}{t}\left\{\left[2\left(\frac{1}{x}-1\right) + \frac{2m^2x}{t}\right]H_1(t) + xG_M^2(t)\right\}\,\,,
\label{elastic}
\end{eqnarray}
where $t = q^2$ is the momentum transfer squared of the photon,
\begin{eqnarray}
 H_1(t) \equiv \frac{G_E^2(t) + \tau G_M^2(t) }{1 + \tau}
\end{eqnarray}
with $\tau \equiv -t/m^2$, $m$ being the nucleon mass, and where $G_E$ and
$G_M$ are the Sachs elastic form factors. In our analysis,  we will use the photon
flux derived in Ref.~\cite{epa}, where an analytical expression is presented.
Moreover, we will also assume that the absorptive factor $S^2_{abs}$, which takes into account of of additional soft interactions between incident protons which leads to an extra production of particles that
destroy the rapidity gaps in the final state \cite{bjorken},  is equal to the unity. Such assumption is a reasonable approximation since the contribution of the soft
interactions is expected to be small in $\gamma \gamma$ interactions due to the long range of the
electromagnetic interaction. { However, it is important to emphasize that the theoretical treatment of $S^2_{abs}$ for $\gamma \gamma$ interactions is still a theme of intense debate in the literature (See, e. g. Refs. \cite{Dyndal:2014yea,Forthomme:2018sxa,Harland-Lang:2020veo,Godunov:2021pdz,Harland-Lang:2021ysd}) . In particular, its dependence on the center - of - mass energy, invariant mass and rapidity of the final state are not well - established, with the predictions for the LHC energy, large invariant masses and central rapidities  being largely distinct, varying between 0.6 and $\approx 1$.  Therefore, our predictions for the production of doubly charged Higgs in $\gamma \gamma$ interactions should be considered an upper bound. Another important aspect is that for exclusive processes mediated by a Pomeron, such factor can be ${\cal{O}}(10^{-2})$ and its precise value is also a theme of intense debate  (for reviews see Refs.~\cite{durham,telaviv,sgap4}).} 

{
Some comments are in order. Firstly, a detailed study of the production of the single charged Higgs $H^{\pm}$ in $pp$ collisions via $\gamma \gamma$ interactions was performed in Ref. \cite{Lebiedowicz:2015cea}, which has demonstrated that the experimental analysis of this process can be useful to probe the existence and properties of $H^{\pm}$. Such study can be considered as complementary to that performed here. In a forthcoming analysis, we intend to improve  Ref. \cite{Lebiedowicz:2015cea} by including the possible $H^{\pm}$ decays  and the full simulation of  backgrounds and pile - up, as considered in the this paper, which will allow us to derive lower mass bounds on $H^{\pm}$. Secondly, in recent years, the determination of photon PDF in a global analysis was performed by different groups assuming distinct assumptions  for, e.g.,  the initial conditions and the treatment of the higher order corrections \cite{Manohar:2017eqh,Bertone:2017bme,Harland-Lang:2019pla}. In general, in these analyzes the authors have considered that the photon PDF is given by the elastic $\gamma^{el}$ and the inelastic $\gamma^{inel}$ contributions (For a more detailed discussion see, e.g. Refs. \cite{daSilveira:2015hha,daSilveira:2021bzs}). The elastic photon distribution  is associated to the probability that a proton emits a photon and remains intact and can be expressed  in terms 
of the  electric and magnetic  form factors  using the Equivalent Photon Approximation (EPA) method \cite{epa,kniehl}, as in Eq. (\ref{elastic}) above.  On the other hand, the inelastic photon distribution $\gamma^{inel}$ provides the probability for a photon emission from a proton in an inelastic interaction and can be estimated assuming that the photon is a constituent of the proton, along with quarks and gluons, with its contribution being  derived by solving the DGLAP evolution equations modified by the inclusion of the QED parton splittings. The resulting predictions for $\gamma^{inel}$, and consequently for the photon PDF,  are  dependent on the assumptions assumed in the global analysis, implying a larger uncertainty on the predictions for the inclusive processes. In contrast, for exclusive (elastic) processes, the uncertainty on the predictions is strongly reduced since it depends only on $\gamma^{el}$, which is well - known.}

\section{Results}
\label{sec:results}

In what follows we will present our results for the doubly charged Higgs production in exclusive processes considereing $pp$ collisions at $\sqrt{s} = 13$ TeV. In our analysis, we will take into account of the current
detector acceptances and resolutions as well as the pile--up effects. We will assume that the $H^{++}H^{--}$ system decays leptonically,
$H^{++}H^{--}\rightarrow e^+e^+ e^-e^-$, both forward protons are tagged and  pile--up is
present. In particular, we will consider two different amounts for the pile-up, $\langle \mu \rangle = 5$ and 50. Moreover, our results will derived by considering additional exclusivity cuts and utilizing the time-of-flight (TOF) detectors for suppressing the combinatorial background coming from pile-up.
The signal is assumed to be the $pp \rightarrow p \otimes H^{++}H^{--} \otimes p  \rightarrow p \otimes  (e^+e^+) (e^-e^-) \otimes p$ process, which will be generated using the   MadGraph~5 \cite{Christensen:2008py, madgraph}. For the background we will consider the photon - induced processes 
$pp \rightarrow p \otimes  e^+e^- e^+e^- \otimes p$ and $pp \rightarrow p \otimes  Z Z \otimes p \rightarrow p \otimes  e^+e^- e^+e^- \otimes p$, where the last process occurs via the box diagram. 
In addition, the inclusive $H^{++}H^{--}$ production via the DY production is also estimated, considering the probably of fake double tagging. All these backgrounds are also generated by MadGraph~5 \cite{madgraph}. As in Refs. \cite{Goncalves:2020saa,Martins:2022dfg}, each event is then properly
mixed with such a number of pile--up interactions that corresponds to the studied
luminosity scenario. Detector effects are incorporated and the pile--up
mixing is done using Delphes~3.5~\cite{delphes3} with input cards with CMS detector specifications.
In order to perform the separation of the  signal from backgrounds, we first select the
central system as in the inclusive processes and then we apply exclusivity criteria. We require both forward protons to be detected by Forward Proton Detectors (FPDs) and processed by Time-of-Flight (ToF) detectors, {  such as the ATLAS Forward Proton detector (AFP)~\cite{afptdr} 
and Precision Proton Spectrometer (CT-PPS)~\cite{Albrow:2014lrm} that are installed symmetrically around the interaction point at a distance of roughly 210~m from the interaction point. The ToF suppression factors are 17.65 and 9.45 for $\langle \mu \rangle =$~5 and 50, respectively, considering that the time resolution of ToF is $\sigma_t =$~10~ps and the signal is collected in a 2-$\sigma_t$ window.  The background with pile-up is evaluated as a combinatorial background coming from the rate of fake double-tagged (DT) events \cite{ToFperformance}.
The final suppression factors representing the rates of fake double-tagged
events for $\langle \mu \rangle$ = 5 and 50 are 0.005 and 0.338, respectively.

}

\begin{figure}[t]
\includegraphics[scale=0.42]{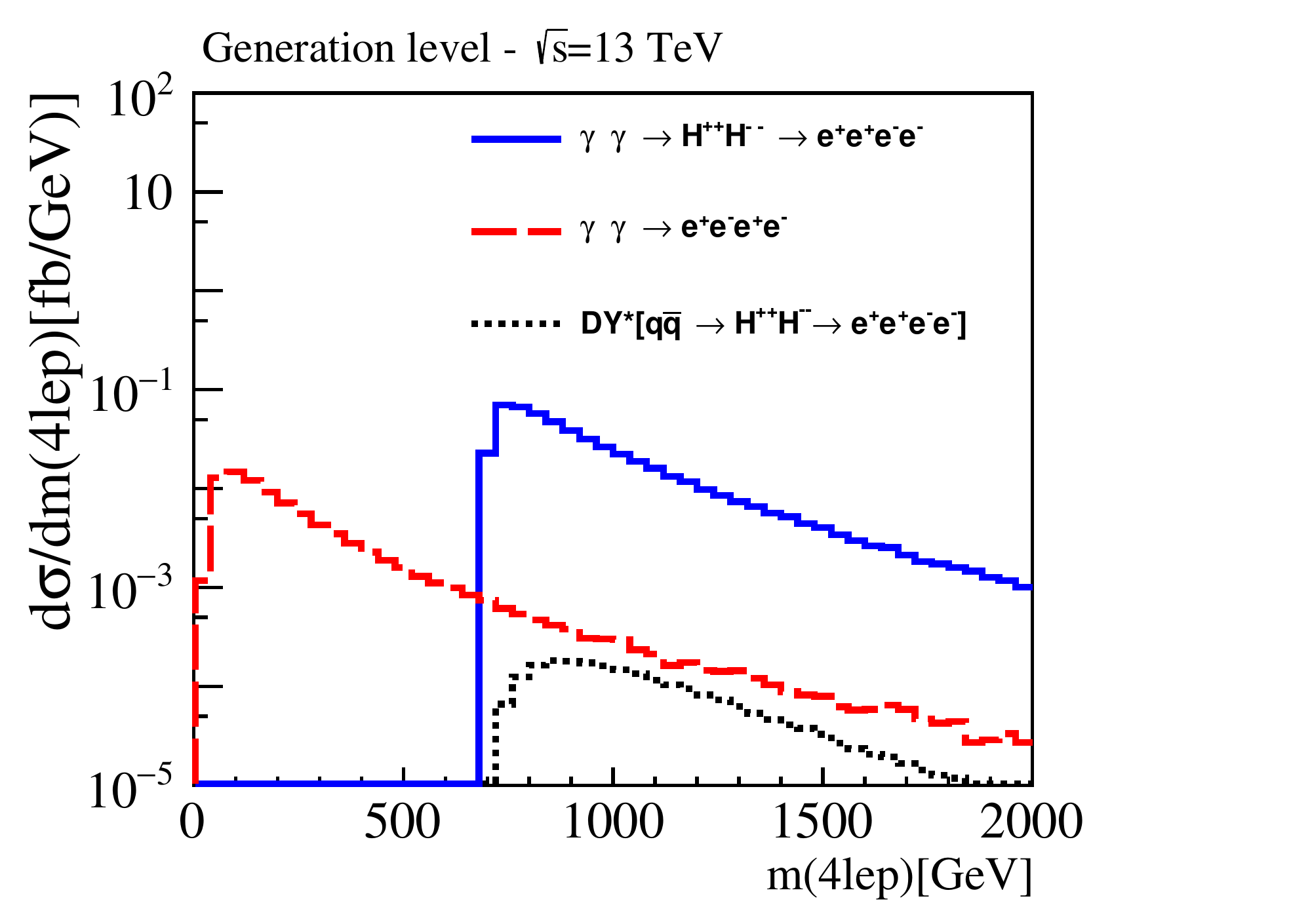} 
\includegraphics[scale=0.42]{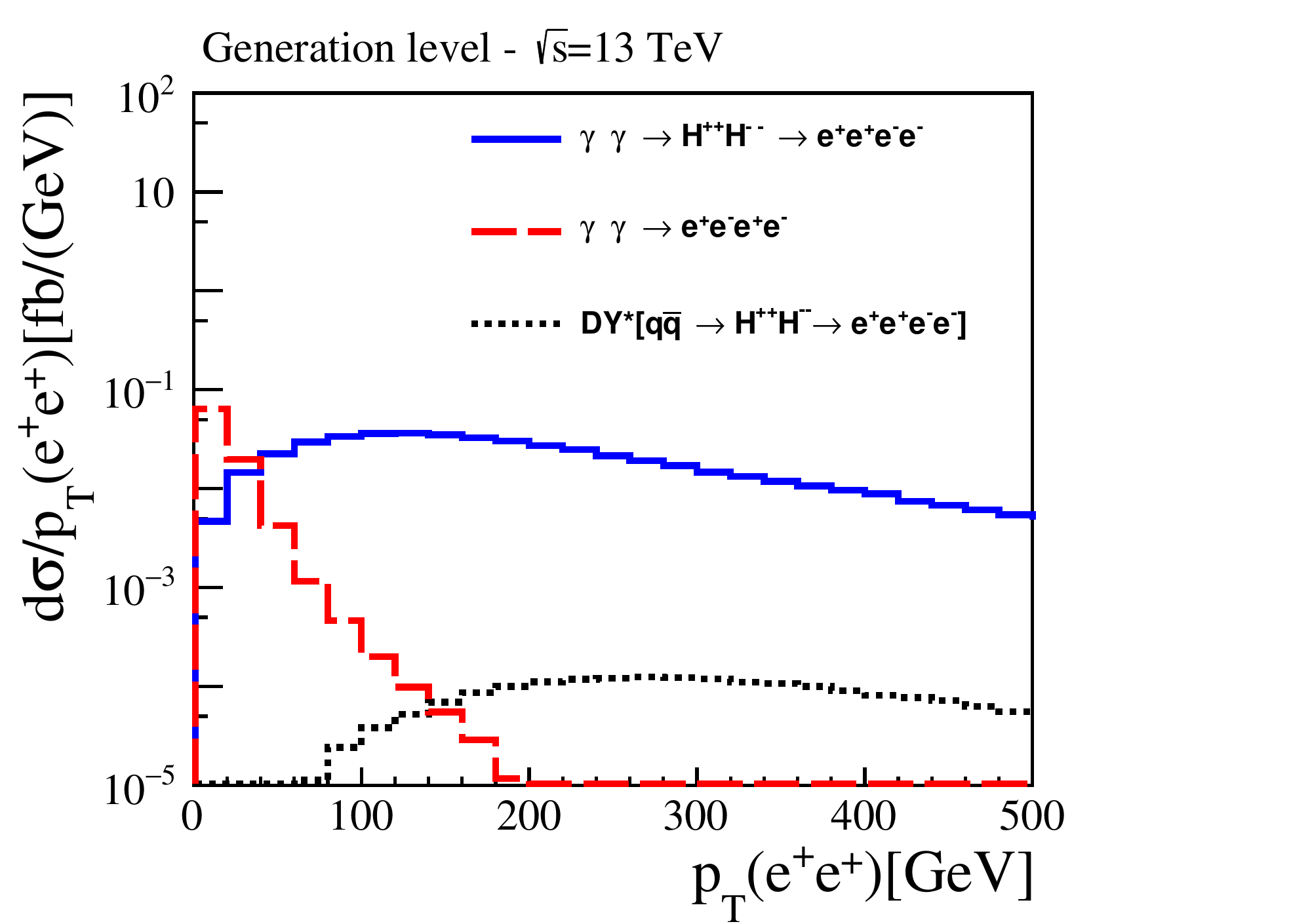} 
\includegraphics[scale=0.42]{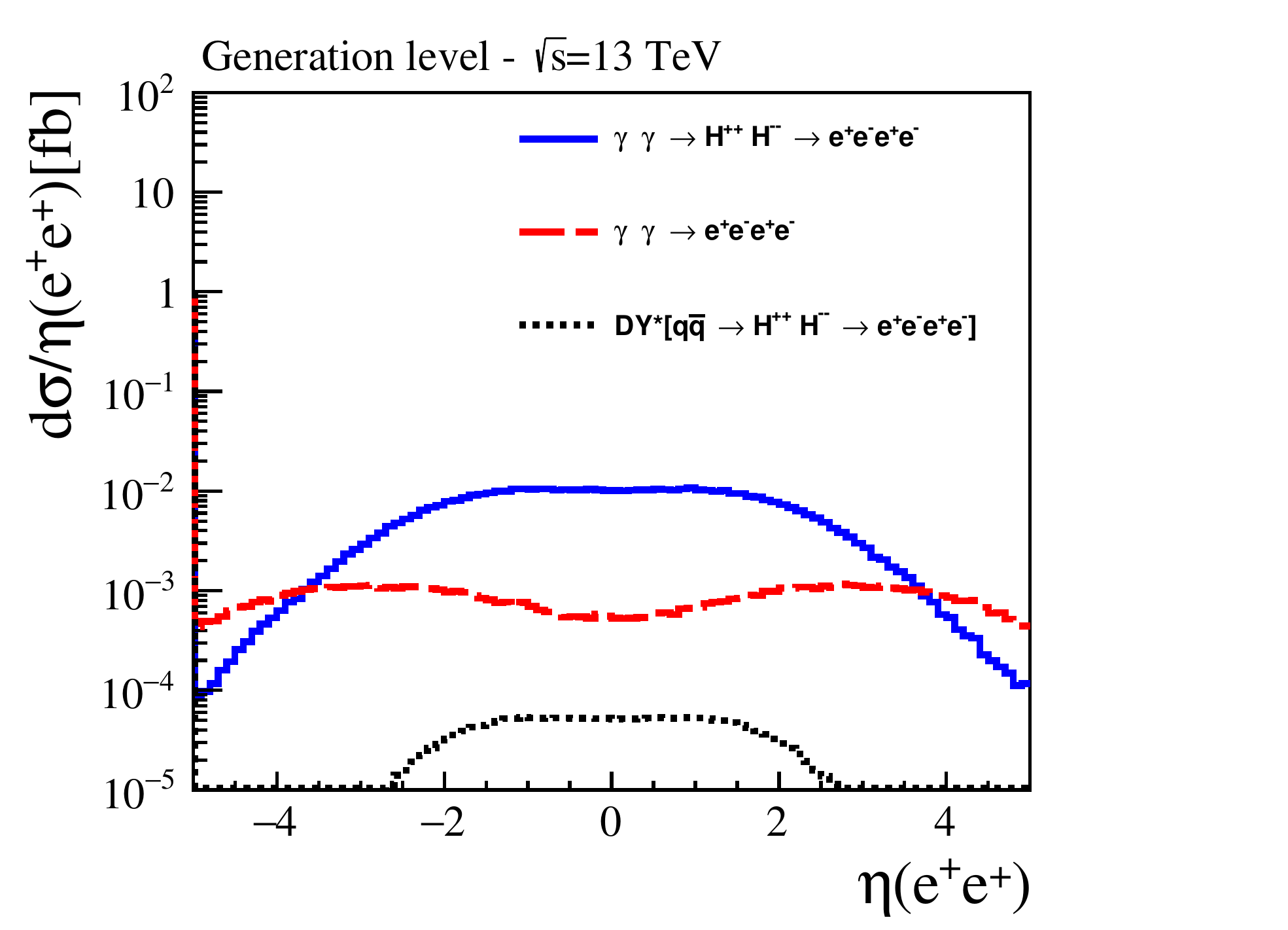} 
\caption{ Predictions for the invariant mass distribution of the four lepton system (upper left panel), transverse momentum (upper right panel) and rapidity (lower panel)  distributions of the $e^+e^+$ pair at the generation level considering $m(H^{\pm\pm}) = 350$ GeV and $pp$ collisions at $\sqrt{s} = 13$ TeV. { DY$^*$ indicates that the double tagging (DT) and time - of - flight (ToF) suppression factors have been applied to the inclusive DY predictions.} }
\label{Fig:pp_13TeV_genplots}
\end{figure}

Initially, we will focus on the $H^{\pm\pm}$ from the Higgs triplet model, in which $Y = 1$, and let's estimate the differential distributions at the generation level. 
In  Fig. \ref{Fig:pp_13TeV_genplots} (upper left panel) we present the associated predictions for the invariant mass distribution of the $e^+e^+ e^-e^-$ system, denoted $m(4lep)$, considering $m(H^{\pm\pm}) = 350$ GeV.
The signal and the main background contributions are shown. The results for the  $pp \rightarrow p \otimes  Z Z \otimes p \rightarrow p \otimes  e^+e^- e^+e^- \otimes p$ process are not presented since they are 6 orders of magnitude smaller than the signal. The background associated to the inclusive DY contribution is evaluated by applying the 
 suppression factors related to the probability of fake double tagging and ToF in the forward detector. As expected, the four lepton production by the QED process $\gamma \gamma \rightarrow e^+e^- e^+e^-$ dominates at small invariant mass, below the production threshold of the doubly charged Higgs pair. On the other hand, for $m(4lep) > 2 m(H^{\pm\pm}) $, one has that this contribution is suppressed by two orders of magnitude in comparison to the signal, but is still larger than the corrected DY background.  

In  Fig. \ref{Fig:pp_13TeV_genplots} (upper right panel) we present our predictions for the transverse momentum distribution of the $e^+e^+$ system, denoted by $p_T(e^+e^+)$. 
One has that the four lepton production by the QED process $\gamma \gamma \rightarrow e^+e^- e^+e^-$ dominates at small $p_T(e^+e^+)$, but is strongly suppressed with the increasing of the transverse momentum. In particular, our results indicate that signal dominates for $p_T(e^+e^+) > 50$ GeV. Moreover, one has that in this kinematical range the signal is larger than the DY contribution by $\approx$ 2 orders of magnitude. { Finally, in Fig. \ref{Fig:pp_13TeV_genplots} (lower panel) we present our predictions for the rapidity distribution of the $e^+e^+$ system, denoted by $\eta(e^+e^+)$. One has that for central rapidities $|\eta(e^+e^+)| \le 2.5$ the signal dominates, with the QED process becoming competitive only for large values of $|\eta(e^+e^+)|$.}

\begin{table}[t]
\begin{center}
\scalebox{0.75}{
\begin{tabular}{|c|c|c|c|c|}
\hline 
{\bf p-p @ $\sqrt{s} = 13$ TeV} & {\bf  $\gamma \gamma \rightarrow H^{++} H^{--} \rightarrow (e^{+} e^{+}) (e^{-}e^{-})$ } & {\bf DY[After DT and ToF]}: & {\bf $\gamma \gamma \rightarrow e^{+} e^{-} e^{+}e^{-}$ }  & {\bf $\gamma \gamma \rightarrow ZZ \rightarrow (e^{+} e^{-})(e^{+}e^{-})$[box]     }            \\
 &  &  $q\bar{q}  \rightarrow H^{++} H^{--} \rightarrow (e^{+} e^{+}) (e^{-}e^{-})$ &     &        \\\hline 
\multicolumn{5}{|c|}{\bf $m(H^{\pm\pm}) = 350.0 $ GeV, $\langle \mu \rangle = 5$ }  \\ \hline
 Total Cross section (fb)                                       & 0.53 &  0.0023  & 0.09   & $\lesssim 10^{-7}$ \\\hline 
p$_{T}(e^{+}e^{+},e^{-}e^{-}) > 50.0 ~{\rm GeV}$            &   0.44    & 0.0023    & $\lesssim 10^{-5} $      &0.0   \\\hline
$  |\eta|<2.5 $ & 0.44 &    0.0023    & $\lesssim 10^{-5} $     & 0.0\\\hline
  $m(H^{\pm\pm})$ $\pm$10 GeV  & 0.34 & 0.0016    &    0.0  & 0.0 \\\hline
  \multicolumn{5}{|c|}{\bf $m(H^{\pm\pm}) = 350.0 $ GeV, $\langle \mu \rangle = 50$ }  \\ \hline
 Total Cross section (fb)                                                    & 0.53 &  0.29  & 0.09   & $\lesssim 10^{-7} $\\\hline 
p$_{T}(e^{+}e^{+},e^{-}e^{-}) > 50.0 ~{\rm GeV}$                          & 0.44   & 0.29    & $\lesssim 10^{-5} $      &0.0   \\\hline
$  |\eta|<2.5 $ & 0.43 &    0.29    & $\lesssim 10^{-5} $     & 0.0\\\hline
  $m(H^{\pm\pm})$ $\pm$10 GeV  & 0.32 & 0.20    &    0.0  & 0.0 \\\hline
   \multicolumn{5}{|c|}{\bf $m(H^{\pm\pm}) = 850.0 $ GeV, $\langle \mu \rangle = 5$ }  \\ \hline
 Total Cross section (fb)                                                    & 0.061 &    0.00013  & 0.09   & $\lesssim 10^{-7} $\\\hline 
p$_{T}(e^{+}e^{+},e^{-}e^{-}) > 50.0 ~{\rm GeV}$                          &  0.060 &   0.00013    & $\lesssim 10^{-5} $      &0.0   \\\hline
$  |\eta|<2.5 $ & 0.060  & 0.00012 & $\lesssim 10^{-5} $     & 0.0\\\hline
  $m(H^{\pm\pm})$ $\pm$10 GeV  & 0.046 & 0.00008    &    0.0  & 0.0 \\\hline
  \multicolumn{5}{|c|}{\bf $m(H^{\pm\pm}) = 850.0 $ GeV, $\langle \mu \rangle = 50$ }  \\ \hline
 Total Cross section (fb)                                                    & 0.061 &  0.016  & 0.09   & $\lesssim 10^{-7} $\\\hline 
p$_{T}(e^{+}e^{+},e^{-}e^{-}) > 50.0 ~{\rm GeV}$                          & 0.060  &   0.016    & $\lesssim 10^{-5} $      &0.0   \\\hline
$  |\eta|<2.5 $ & 0.060 &    0.016    & $\lesssim 10^{-5} $     & 0.0\\\hline
  $m(H^{\pm\pm})$ $\pm$10 GeV  & 0.046 & 0.010    &    0.0  & 0.0 \\\hline
\end{tabular}}
\caption{Predictions for the total cross sections of the signal and background at the detector level considering two different values of the doubly charged Higgs mass and two scenarios for the pile-up. { The model considered here is the Higgs triplet model ($Y = 1$).}}
\label{tab:cuts_visible_xs_delphes}
\end{center}
\end{table}

In Table \ref{tab:cuts_visible_xs_delphes} we present our predictions at the detector level, which have been derived assuming the following cuts: (a) $p_T(e^{\pm} e^{\pm}) > 50$ GeV; (b) that the leptons are in the rapidity range $|\eta| \le 2.5$, and (c) that the invariant mass of the dilepton system $m(e^{\pm} e^{\pm})$ is in the range  $m(H^{\pm\pm})$ $\pm$ 10 GeV. The results are presented for two values of the doubly charged Higgs and two scenarios for the pile-up. One has that the signal and corrected DY cross sections, denoted DY[After DT and ToF] in the Table, decrease for larger values of $m(H^{\pm\pm})$, with the inclusive DY production being the main background. For a small amount of the pile-up and $m(H^{\pm\pm}) = 350$ GeV, one has that signal is two orders of magnitude larger than the background, as expected from Fig. \ref{Fig:pp_13TeV_genplots}.   Such conclusion is also valid for $m(H^{\pm\pm}) = 850$ GeV and $\langle \mu \rangle = 5$. In contrast, for a larger amount of pile-up, $\langle \mu \rangle = 50$, the difference between signal and background decreases and both cross sections become of the same order of magnitude, with the signal being still larger than the corrected DY prediction.

\begin{figure}[t]
\includegraphics[scale=0.41]{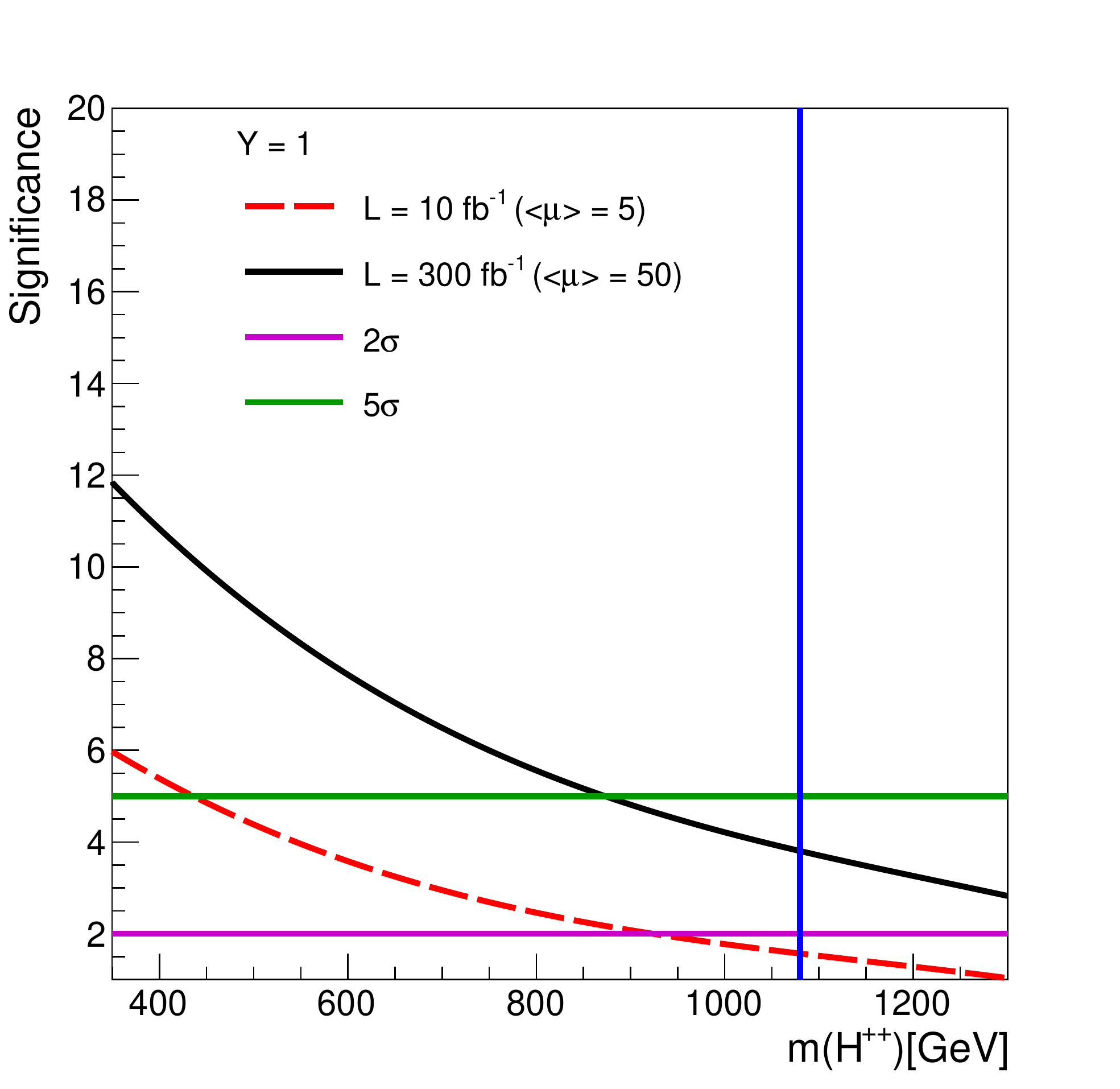} 
\includegraphics[scale=0.41]{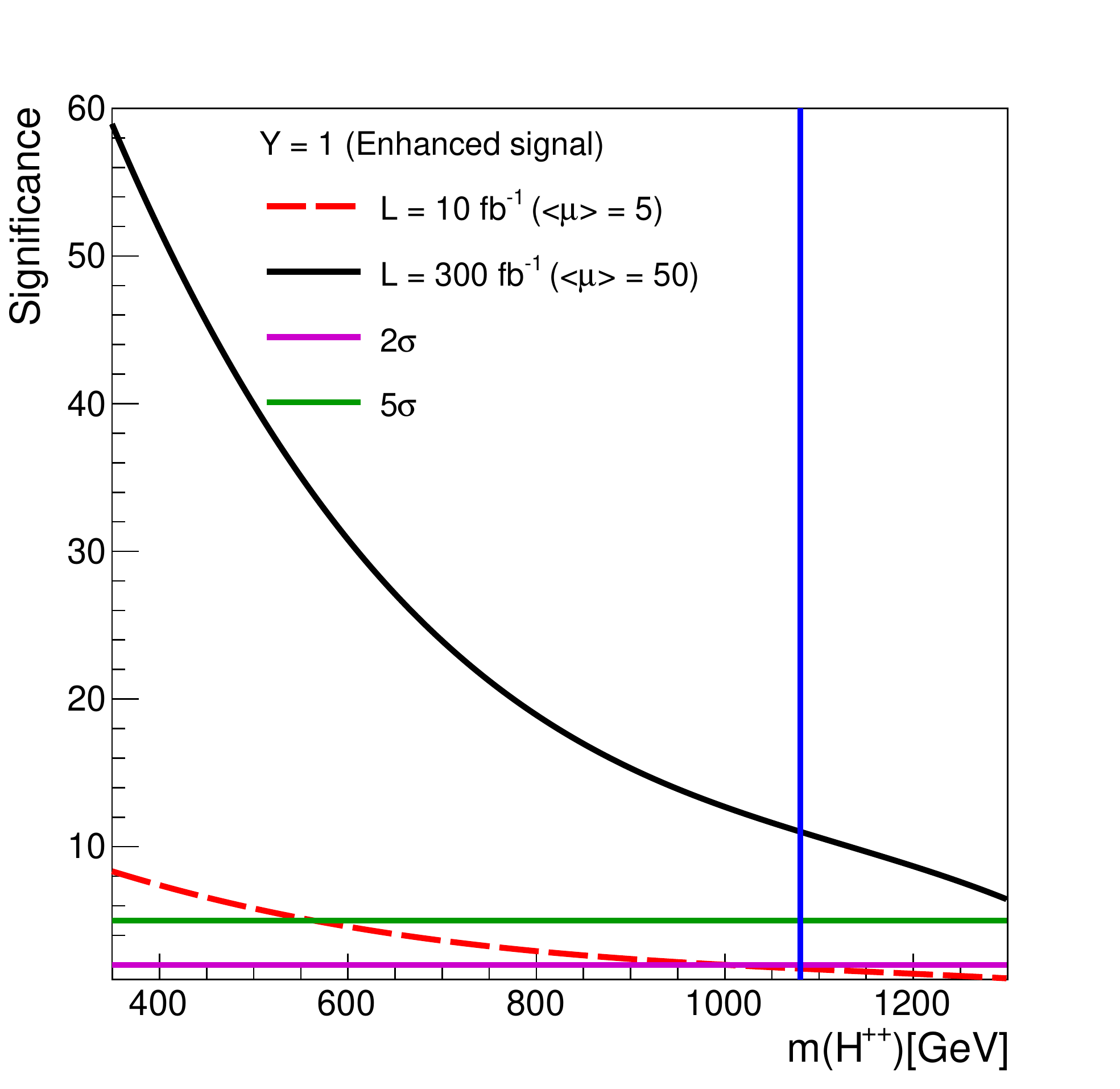} 
\caption{{ Mass dependence of the statistical significance  for the exclusive doubly charge Higgs production in $pp$ collisions at $\sqrt{s} = 13$ TeV considering two scenarios for the pile-up configuration and integrated luminosity $(<\mu>,L)$ and to distinct scenarios for the background: (a) the standard approach considering the inclusive DY production as a background (left panel), and (b) that the remaining contribution of the inclusive process enhance the number of doubly charged Higgs events (right panel). Results for the Higgs triplet model  ($Y = 1$).  The vertical black line indicates the current upper value for  $m(H^{\pm\pm}) = 1080$ GeV obtained in Ref.  \cite{ATLAS:2022pbd}.}}
\label{Fig:stat_sig_y1}
\end{figure}

The previous results allows to estimate  the statistical significance and the signal to background ratio, $S/B$. In order to estimate these quantities, we consider
two luminosity scenarios in terms of $\langle \mu \rangle$ and $\cal L$, where
$\langle \mu \rangle$ represents the average number of pile-up interactions per event  and $\cal L$ is the integrated luminosity. Moreover, we assume $\cal L$ to be 10 and
300~fb$^{-1}$ for $\langle \mu \rangle =$~5 and 50, respectively. In addition, the significance will be estimated  using the formula based on Asimov data set~\cite{Cowan:2010js}, which is more reliable for $S \approx B$ and   reduces to the $S/\sqrt{B}$ ratio  if $S \ll B$. The results are presented in 
Fig. \ref{Fig:stat_sig_y1} (left panel) as a function of  doubly charged Higgs mass. 
{We restrict our study for the mass range $m _{H^{\pm\pm}} \lesssim 1300$ GeV due to the current limitations of the forward detectors, which are not able to measure final states with an invariant mass larger than $\approx 2600$ GeV, due to the limited detector angular acceptance for the outgoing protons.}
{Our results indicate that the ATLAS and CMS forward detectors have the potential to detect a $H ^{\pm\pm}$ particle with a significance higher than $5 \sigma$, if it has a mass up to $\sim 870$ GeV, and hint at the existence of such a particle, with a significance of at least $3 \sigma$, throughout the entire mass range from very small masses up to the limit of $m _{H ^{++}} \sim 1300$ GeV, set by the detector acceptance. In the mass window between the $870$ GeV and $1300$ GeV, however, the $H ^{++}$ detection would require an independent confirmation from other experiments, possibly from inclusive searches, 
to cross the $5 \sigma$ threshold and claim a discovery.
In any event, it is clear that a future experimental analysis of the exclusive production at the HL-LHC will be able to improve the current limits on $m _{H ^{\pm\pm}}$, obtained by the analysis of inclusive processes which, for an $H ^{++}$ belonging to a scalar triplet, { are around $1080$ GeV} \cite{ATLAS:2022pbd}.}

{ 
In the previous analysis we have considered the standard approach  that the contribution of inclusive processes is a background for the exclusive production. However, as our goal is to probe the existence of the doubly charged Higgs and determine its mass, the remaining inclusive DY events after the implementation of the exclusive cuts can also be considered as  signals of the  type
II seesaw mechanism. As a consequence, the signal becomes the sum of the exclusive and DY[After DT and TOF] contributions and  the statistical significance is enhanced. The associated predictions for this scenario are presented in Fig. \ref{Fig:stat_sig_y1} (right panel). Our results indicate that, in this scenario,  the significance is larger than  $5 \sigma$ for $\langle \mu \rangle = 50$ in the range not yet covered by the inclusive measurements. In what follows, we will restrict our analysis to the standard procedure, but the reader should keep in mind that the predictions for the significance must be considered a pessimistic estimate and represent a lower bound for the number of events that are predicted for the future runs of LHC.}

\begin{figure}[t]
\includegraphics[scale=0.42]{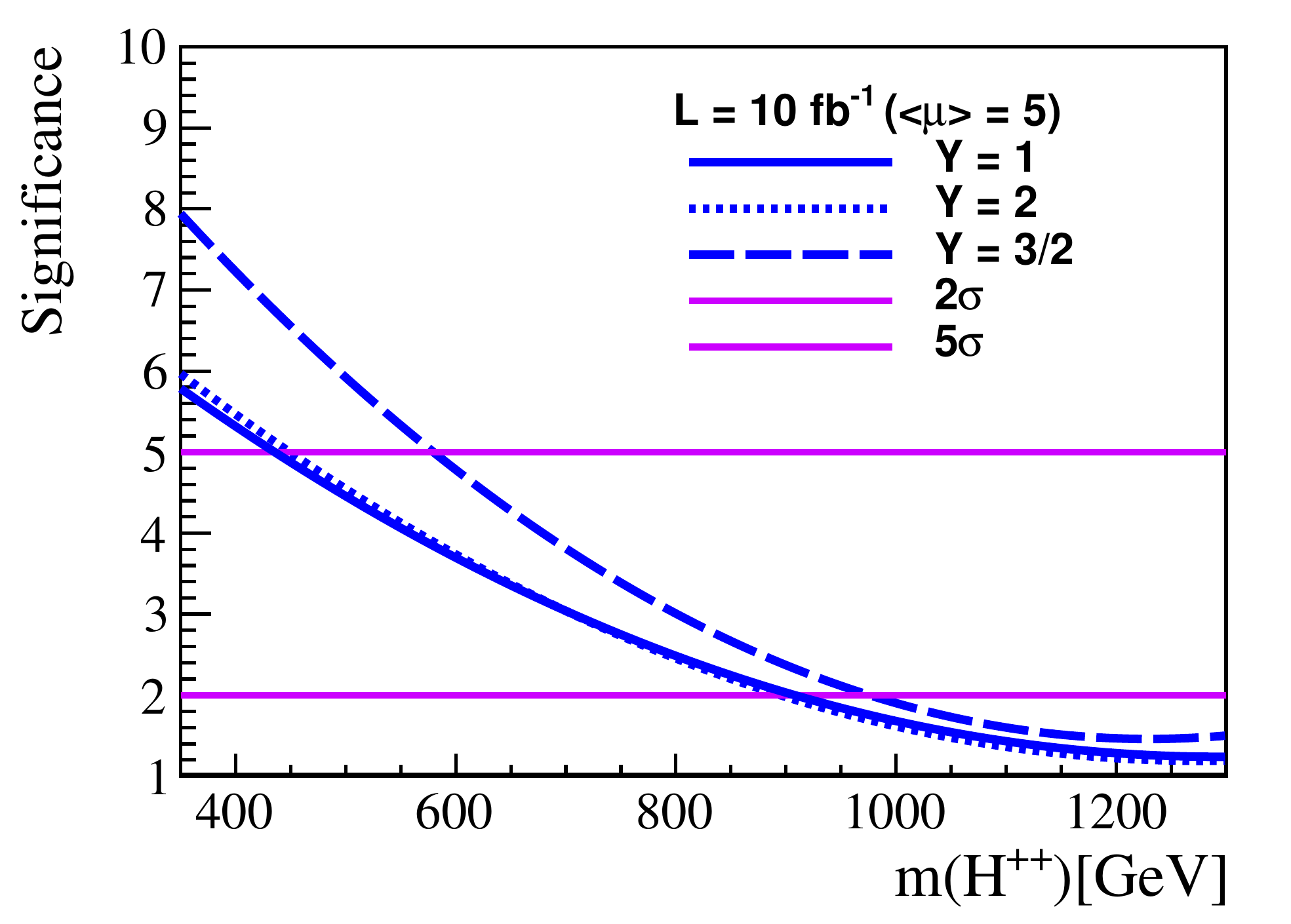}
\includegraphics[scale=0.42]{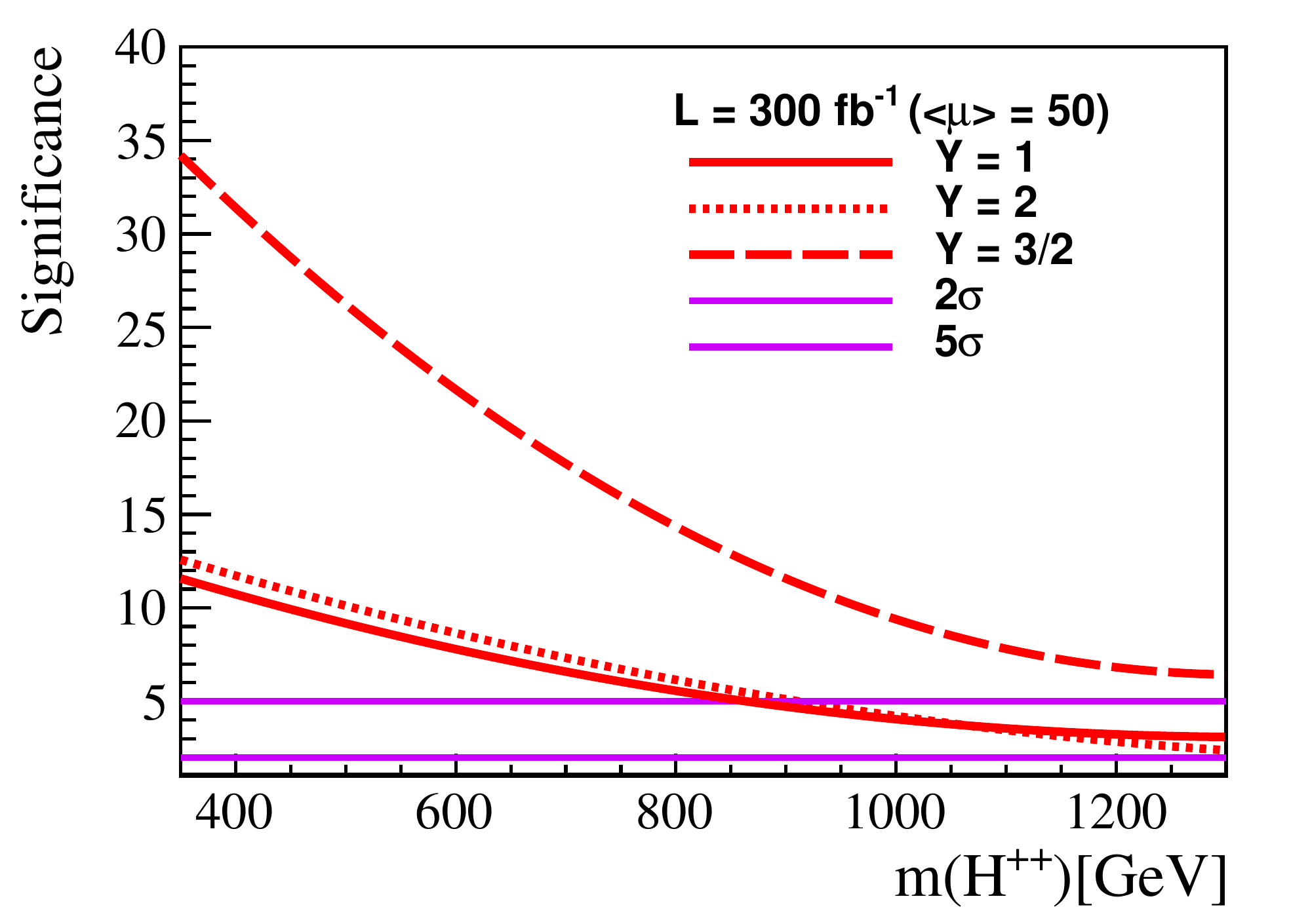}
\caption{Mass dependence of the statistical significance  for the exclusive doubly charge Higgs production in $pp$ collisions at $\sqrt{s} = 13$ TeV considering distinct $SU(2) _L$ representations. and different scenarios for the pile-up configuration and integrated luminosity $(<\mu>,L)$.}
\label{Fig:stat_sig_difsy}
\end{figure}

{In Fig. \ref{Fig:stat_sig_difsy} we present the results of the standard analysis applied to the cases where the $H^{\pm\pm}$ particle arises from more general $SU(2) _L$ representations, as discussed in the Section \ref{sec:form}. Besides the triplet representation ($Y = 1$) we also analyzed the singlet ($Y = 2$) and doublet ($Y = 3/2$) cases.
It is apparent from the plots that the discovery potential for the singlet scenario is very similar to that of the triplet case, discussed above. On the other hand, one can notice a substantial enhancement on the sensitivity for the case of the doublet representation, which enjoys a detection prospect above $5 \sigma$ in the entire mass range up to $1300$ GeV. This result can be understood by looking at the behavior of the $H ^{++}$ production cross section via DY and $\gamma \gamma$ fusion processes. Notice that while the $\gamma \gamma$ fusion is indifferent to the multiplet the $H ^{++}$ originates from, the DY process depends on the strength of the $Z _\mu H ^{++} \partial ^\mu H ^{--}$ coupling, which varies for the different multiplets. According to the Eq. \eqref{eq:lag_kin}, we have $\mathcal{C} _{Z H ^{++} H ^{--}} = \dfrac{g}{c _W} (2 - Y - 2 s _W ^2)$, such that: $\mathcal{C} _{Z H ^{++} H ^{--}} = -0.388$ for the singlet; $\mathcal{C} _{Z H ^{++} H ^{--}} = 0.034$ for the doublet and $\mathcal{C} _{Z H ^{++} H ^{--}} = 0.455$ for the triplet. Notice that while the singlet and triplet couplings have similar magnitudes (with opposite signs that, nevertheless, does not affect the final cross section), the coupling in the doublet case is one order of magnitude smaller, leading to a smaller DY production cross section of the $H ^{++}$. Therefore, an even larger suppression of the DY background takes place in the doublet case, as compared to the singlet and triplet cases, thus enhancing the sensitivity. 
Notice further that, while this small coupling increases the sensitivity of the exclusive search, it operates in the opposite direction for the traditional inclusive searches, insofar as it reduces the number of $H ^{++}$ signal events.  
Therefore, in this particular case of the $H ^{++}$ from a scalar doublet, the exclusive search using the forward LHC detectors features a clear advantage compared to the inclusive searches focusing on the central detectors only.
As for the other cases, in which the exclusive analysis has a comparable sensitivity compared to the inclusive searches, it is important to emphasize the complementary aspect of this kind of analysis, which has the potential to strengthen the results obtained with the usual inclusive processes, be it a constraint or a discovery.
}

\section{Summary}
\label{sec:sum}

 One of the challenge problems in Particle Physics today is the description of the neutrino masses and all neutrino oscillation data. One of the simplest ways to describe these data is provided by the type II seesaw model, which predicts the existence of seven physical Higgs bosons, two of them being doubly charged.
Another feature of the type II seesaw model is that the Higgs triplet can be produced directly through gauge interactions with electroweak bosons. Such aspect has motivated the study of the doubly charged Higgs pair production through the Drell - Yan process, where the pair is produced via $s$ - channel $\gamma^*/Z$ exchange, or via the vector boson fusion processes $\gamma \gamma \rightarrow H^{++} H^{--} $ and $Z Z \rightarrow H^{++} H^{--} $. In recent years, several experimental studies in high - energy colliders have searched for  doubly charged scalar particles and derived bounds on the $H^{\pm\pm}$ mass. Distinctly from previous studies, which focused on inclusive reactions, in this paper we have proposed the study of exclusive processes as an alternative to searching for the doubly charged Higgs. As demonstrated in our analysis, although the associated cross sections are smaller than those predicted for the inclusive reactions, the very clean final state and possibility of tagging of the intact protons in the final states and the reduction of the impact of the pile-up using the time - of - flight detectors, makes feasible the study of the exclusive  $H^{++} H^{--}$ production in $pp$ collisions at the LHC. 
{In particular, we have shown that the ATLAS and CMS forward detectors can extend the $H ^{++}$ searches towards higher masses, complementing and improving the current limits obtained by the inclusive processes. }

\begin{acknowledgments}
V.P.G. and D.E.M. acknowledge useful discussions with M. Tasevsky and G. Gil da Silveira. This work was partially supported by INCT-FNA (Process No. 464898/2014-5) and Simons Foundation (Award
Number:884966, AF). V.P.G. was partially supported by the CAS President's International Fellowship Initiative (Grant No. 2021VMA0019) and by CNPq, CAPES and FAPERGS. T.B.M. acknowledges CNPq for the financial support (grant no. 164968/2020-2). D.E.M. was partially supported by CNPq-UFPel (grant no.164609/2020-2 and 152473/2022-0) along with The Henryk Niewodniczanski Institute of Nuclear Physics Polish Academy of Sciences (grant no. UMO-2021/43/P/ST2/02279). FSQ thanks Serrapilheira Foundation grant Serra-1912-31613,
Simons Foundation (Award Number:884966, AF), 
FAPESP grant 2021/01089-1,FAPESP grants 2021/14335-0, CNPq grant 307130/2021-5, and ANID-Programa Milenio-code $ICN2019\_044$.
\end{acknowledgments}


\end{document}